\documentclass[12pt,a4paper]{article}
\usepackage{amsthm}
\theoremstyle{plain}
\usepackage{authblk}
\usepackage[english]{babel}
\usepackage{graphicx}
\usepackage[latin1]{inputenc}
\usepackage{verbatim}
\usepackage{amsfonts}
\usepackage{amsmath}
\usepackage{txfonts}
\usepackage[T1]{fontenc}
\usepackage{color}
\usepackage{epsfig}
\usepackage{natbib}
\usepackage{url}
\usepackage[bookmarksnumbered=true, colorlinks=true, citecolor=blue]{hyperref}
\date{}

\title{On the Importance of Interpretation in Quantum Physics. A Reply to Elise Crull.}
\author{Antonio Vassallo}
\author{Michael Esfeld}
\affil{University of Lausanne, Department of Philosophy, CH-1015 Lausanne} 

\begin{document}

\maketitle
\begin{center}
Accepted for publication in \emph{Foundations of Physics}.
\end{center}
\pdfbookmark[1]{Abstract}{abstract}
\begin{abstract}
\citet{416} claims that by invoking decoherence it is possible (i) to obviate many ``fine grained'' issues often conflated under the common designation of measurement problem, and (ii) to make substantial progresses in the fields of quantum gravity and quantum cosmology, without any early incorporation of a particular interpretation in the quantum formalism. We point out that Crull is mistaken about decoherence and tacitly assumes some kind of interpretation of the quantum formalism.
\end{abstract}

In the article ``Less Interpretation and More Decoherence in Quantum Gravity and Inflationary Cosmology'', Elise Crull praises the virtues of decoherence in addressing many conceptual problems in quantum physics -- including the fields of quantum gravity and quantum cosmology, and argues that decoherence does such a good job \emph{alone}, that is, without drawing from the outset upon any interpretational work. While we agree with the author that decoherence is a powerful \emph{tool} that effectively helps us dealing with the conceptual asperities of quantum physics -- the measurement problem above all --, we do not see how this fact might dispense us with the urge to interpret the bare quantum formalism.\\
Let us start from what, in our opinion, represents the apple of discord here: the notion of interpretation. What does Crull intend by this word? In the paper, she does not give a straightforward definition of what interpretation is to her. Presumably, she considers de Broglie-Bohm and Everett approaches, plus collapse theories, as instances of interpretation. This set of choices is rather heterogeneous, ranging from frameworks that simply add some extra-physical assumption on top of the quantum formalism (e.g. the existence of many worlds or many minds), to approaches that modify (or supplement) the usual formulation of quantum dynamics (e.g. by adding a ``guiding'' equation to the usual Schr\"odinger one). What is the conceptual need that all these frameworks attempt to satisfy? The answer seems crystal clear to us: to find a bridge that connects the quantum formalism to the physical world (or with the ``impression'' of there being a physical world). And, setting aside instrumentalist inclinations, the only reasonable way to establish such a link is by specifying an ontology for the theory.\footnote{Of course, such a specification should not necessarily amount to bringing ``classical terms'' into the equations, as suggested most notably by \cite{415}, but it could include the commitment to irreducibly ``non-classical'' features of reality: Everett's interpretation is a remarkable example, in this sense.} If we agree that this is a cogent -- although generic -- characterization of what ``interpreting'' quantum physics stands for, we immediately see that the need for an interpretation does not arise just when we get to explain the behavior of a macroscopic measurement apparatus, but it comes to light as soon as we ask ``what is the picture of the physical world conveyed by the theory?''. Under this reading, to say that <<decoherence [...] brings with it no supplement to the physics [...] It introduces no new physical principles, nor is it an interpretation>> (\citealp{416}, p. 11) is misleading: decoherence and interpretation belong to different conceptual categories, and only an interpretation can be predicated of ``bringing no supplement to physics'', or ``introducing no new physical principles''. In short, by claiming that decoherence can do a lot without interpretation, Crull is already presupposing a fair amount of interpretation.\\
To see why decoherence needs an interpretation from the outset, let us consider, for simplicity's sake, the standard Schr\"odinger's cat experimental setting. As the author herself acknowledges (ibid., p. 15), the global decoherence process remains entirely within the unitary Schr\"odinger evolution. Consequently, with whatever system -- including whatever environment -- the cat interacts, this interaction will simply lead to more entanglement. Even if the standard is reduced from accounting for a definite state of the cat to accounting for why the cat appears to be in a definite state, decoherence as such does not do the job. Consider an observer being coupled to the cat. Then the whole system will go into an entangled state that is a superposition of ``cat alive and live cat appearing to observer'' and ``cat dead and dead cat appearing to observer''. In brief, because decoherence by no means touches the fact that the state of the whole system is such a superposition instead of one in which the observer is in a state of having definite appearances, decoherence without an interpretation cannot even account for the appearance of there being physical systems in definite states to observers.\\
By contrast, Crull claims that decoherence alone explains why (a) the cat is observed in a position eigenstate rather than, say, an energy one (problem of the preferred basis), (b) the observation of, say, the ``dead'' part of the cat state is not influenced by  the ``alive'' part of the cat state (problem of the non-observability of interference), and (c) the cat is observed either alive or dead (generic problem of outcomes). As a matter of fact, we do not need to refer to observations in a standard sense, here: if we agree that, once the cat state gets entangled with the environment, the physical information encoded in the former is transferred in the latter, then the above assertions can be cast in terms of environmental degrees of freedom ``recording'' information about the cat qua quantum system.  Even with this further conceptual sophistication, our argument stands firm: if Crull wants the decoherence mechanism to explain the assertions (a)-(c) she needs to make a minimal set of assumptions, e.g., that (i) the cat is a quantum system completely described by its own individual state - at least, before the decoherence process takes place, (ii) there is something distinct from the cat, which we call ``environment'' - again, at least before decoherence, (iii) the physical information transfer happens in virtue of the fact that (the states of) the cat and the environment stand in a real physical relation we call ``entanglement'',\footnote{This particular assumption is not as uncontroversial as it might seem. A quantum-Bayesianist, for example, would tend to deny it.} and that (iv) such an information transfer gives rise to a state of affairs which is ``non-classical'' (e.g. there not being a fact of the matter about whether the cat is definitely alive or dead). If we explicitly drop (i)-(iv), we are left with no explanans whatsoever. Just to be clear: we are not claiming that (i)-(iv) is the \emph{only} possible interpretation compatible with the decoherence mechanism,\footnote{This is why we agree with the author that it is not the case that decoherence can be understood only in an Everettian framework.} but that decoherence needs an interpretational background in order to make physical sense at all.\\
To make a concrete case, consider the claim that <<[...] by pursuing the consequences of quantum principles, we learn that it is a quantum feature of the world - namely, the universality of entanglement and the decoherence processes resulting therefrom - that gives rise to states of affairs that are empirically indistinguishable from ``classical'' states of affairs>> (ibid., pp. 12-13). It is now clear that such a claim hides a huge amount of interpretational work: there is, in fact, no necessary connection between the universality of entanglement (and the resulting decoherence) and the rise of ``non-classical'' states of affairs. Evidently, to claim that the universality of entanglement gives rise to ``non-classical'' or also ``semi-classical'' states of affairs is pursuing the consequences of a certain interpretation. Nothing prevents us from adopting a different interpretive framework such as, for example, the formulation of the de Broglie-Bohm theory put forward in \citet{323}: also this framework takes the universality of entanglement very seriously -- and accommodates decoherence processes as well. Indeed, decoherence acquires a precise meaning in this context, and represents a key ingredient in the account of the transition from quantum to classical trajectories of particles. But decoherence can accomplish this task only against the background of a primitive ontology of there being always a definite configuration of particles in spacetime. In other words, decoherence is important to understand the link between the quantum formalism and the classical world, but only on the basis of having settled for an ontology of quantum physics in the first place.
To sum up, we claim that Crull did not make a case in favor of decoherence and against interpretation, but just defended a particular brand of interpretation against others, based on the alleged fact that the one proposed by her is the most parsimonious. Crull, despite her claims on the contrary, is yet another gladiator battling in the interpretational arena of quantum physics.\\ 
In conclusion, the moral we want to convey is simply that decoherence per se does not do anything when it comes to establishing a connection between the quantum formalism and the physical world (e.g. by accounting for definite states of physical systems or the appearance of such states to observers), but decoherence integrated in an interpretive framework does a good job to accomplish this task. This is because it is the interpretive framework that sets the explanantes in which decoherence enters, and hence decoherence alone cannot be said to \emph{explain} anything in a physically interesting sense. Once this fact is recognized, we totally side with Crull in claiming that the quest for a quantum theory of gravity might immensely benefit from the inclusion of decoherence mechanisms into it.

\pdfbookmark[1]{Acknowledgements}{acknowledgements}
\begin{center}
\textbf{Acknowledgements}:
\end{center}
Antonio Vassallo acknowledges support from the Swiss National Science Foundation, grant no. $105212\_149650$.

\pdfbookmark[1]{References}{references}
\bibliography{biblio}
\end{document}